\documentclass[12pt]{article}
\usepackage{amsmath,amssymb}
\usepackage{layout}
\usepackage{latexsym,graphicx}
\usepackage[cp1251]{inputenc}
\usepackage{curves,eepic}

\usepackage{floatflt}

\usepackage{wrapfig}
\usepackage{amsmath,amssymb}

\newcommand{\be}{\begin{equation}}
\newcommand{\ee}{\end{equation}}
\newcommand{\bes}{\begin{subequations}}
\newcommand{\ees}{\end{subequations}}
\newcommand{\bea}{\begin{eqnarray}}
\newcommand{\eea}{\end{eqnarray}}
\newcommand{\bear}{\begin{equation}\begin{array}}
\newcommand{\eear}[1]{\end{array}\label{#1}\end{equation}}
\def\ba{$$\begin{array}}
\def\ea{\end{array}$$}
\def\bra{$\begin{array}}
 \def\era{\end{array}$}

\newcommand{\bm}{\boldmath}
\newcommand{\fr}[2]{\dfrac{{ #1}}{{ #2}}}

\newcommand{\la}{\langle}
\newcommand{\ra}{\rangle}

\def\vep{{\varepsilon}}

\newsavebox{\fmbox}

{\end{list}}
\newcounter{enumct}

\newcommand{\bu}{$\bullet$\ }

\begin{document}
\date{}

\title{Neutrino Factory Based on
Linear Collider}

\author{ I.~F.~Ginzburg\\
{\small\it Sobolev Institute of Mathematics}\\ {\small\it
and Novosibirsk State University,}\\ {\small\it
Novosibirsk, 630090, Russia}}

\maketitle

\begin{abstract}
The beams of Linear Collider after main collision can be
utilized to build neutrino factory with exceptional
parameters. We also discuss briefly possible applications
of some elements of proposed scheme for standard fixed
target experiments, new experiments with $\nu_\mu N$
interactions and in material sciences.
\end{abstract}


\section{Introduction}

The project of Linear Collider (LC) contains one essential
element that is not present in other colliders. Here each
electron (or positron or photon) bunch will be used only
once, and physical collision leave two very dense and
strongly collimated  beams of high energy electrons or/and
photons with precisely known time structure. We consider,
for definiteness, electron beam parameters of the TESLA
project \cite{TESLA}
  \bear{c}
particle\;\; energy\;\; E_e=250\;GeV, \\ number\;\;
of\;\;electrons\;\; per\;\; second\;\; N_e=2.7\cdot
10^{14}/s,\\ mean\;\;beam\;\; power\;\;P_b\approx 11\;MWt,\\
transverse\;\; size\;\; and\;\; angular\;\;
spread\;\;negligible.
 \eear{beampar}

The problem, how to deal with this powerful beam dump, is
under intensive discussion.

Main discussed variant is to destruct these used beams with
minimal radioactive pollution (see e.~g.~\cite{TESLA}). It
looks natural also to use these once--used beams for fixed
target experiments with unprecedented precision.

Recently we suggested to utilize these used beams  to
initiate work of subcritical fission reactor and to
construct neutrino factory \cite{LCWS05}. Here we present
estimates for one of these options. Real choice and
optimization of parameters should be the subject of detail
subsequent studies.\\

\bu The study of neutrino oscillations is one of the most
important problems in modern particle physics. In this
problem the neutrino factories promise most detailed and
important results. The existing projects of neutrino
factories (see e.g. \cite{nufact,nufact1}) are very
expensive and their physical potential is limited by
expected neutrino energy and productivity of neutrino
source.

The neutrino factory based on LC is much less expensive
than those discussed nowadays \cite{nufact,nufact1}. The
combination of a high number of particles in the beam and
high particle energy \eqref{beampar} provides very
favorable properties of neutrino factory. The initial beam
will be prepared in LC irrelevantly to the neutrino factory
construction. The construction demands no special
electronics except for that for detectors. The initial beam
is very well collimated so that the additional efforts for
beam cooling are not necessary. The use of the Ice-cub in
Antarctic as a far distance detector (FDD) allows to see
possible oscillations $\nu_\mu\to sterile\; \nu$ via
measurement of deficit of $\nu_\mu N\to \mu X$ events.

The neutrino beam will have very well known discrete time
structure  that repeats the same structure in the LC. This
fact allows to separate cosmic and similar backgrounds with
high precision during operations. Very simple structure of
neutrino generator allows to calculate the energy spectrum
and content of the main neutrino beam with high accuracy.
It must be verified with high precision in nearby detector
(NBD).

In this project neutrino beam will contain mainly muon
neutrino's and antineutrino's with small admixture $\nu_e$
and $\bar{\nu}_e$ and tiny dope of $\nu_\tau$ and
$\bar{\nu}_\tau$ (the latter can be calculated with low
precision). The neutrino energies are spread up to about
80~GeV with mean energy about 30 GeV, providing reliable
observation of $\tau$, produced by $\nu_\tau$ from
$\nu_\mu-\nu_\tau$ oscillations. In the physical program of
discussed $\nu$ factory we consider only problem of
oscillations $\nu_\mu-\nu_\tau$ and/or $\nu_\mu
-\;sterile\;\nu$. The potential of this $\nu$ factory in
other problems of $\nu$ physics should be studied after
detailed consideration of the project.\vspace{-3mm}

\section{Elements of neutrino factory}

\subsection{Scheme}

The proposed scheme deals with the electron beam used in LC
and contains the following parts (see Fig.~1).
\begin{figure}[hbt]
 \includegraphics[width=0.95\textwidth, height=2.5cm]{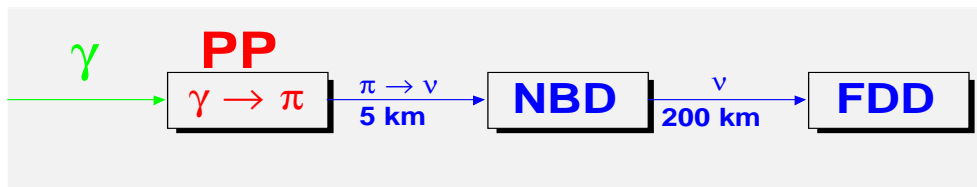}
\caption{Main parts of neutrino factory.}
 \end{figure}

\centerline{\bu Pion producer (PP), \, \, \bu Neutrino
transformer (NT),\, \, \bu Nearby detector (NBD),}

\centerline{\bu Far distance detector (FDD), \, \, \bu Beam
turning magnet (BM) before PP.}

\subsection{Beam turning magnet}

The system should start with the magnetic system which
turns the used beam at an angle necessary to reach FDD with
sacrifice of monochromaticity but without growth of angular
spread. The vertical component of turning angle $\alpha_V$
is determined by Earth curvature. Let us denote the
distance from LC to FDD by $L_F$. To reach FDD the initial
beam (and therefore NT) should be turned before PP at the
angle $\alpha_V =\arcsin [L_F/(2R_E)]$ below horizon  (here
$R_E$ is Earth radius).

The horizontal component of turning angle can be minimized
by suitable choice of the proper LC orientation
(orientation of incident beam near the LC collision point).

\subsection{Pion producer (PP)}

The next stage is pion production in the PP in the form of
a 20~cm long water cylinder ({\it 20~cm is one radiation
length}). The water in cylinder should rotate for cooling.
In this PP almost each electron will produce bremsstrahlung
photon with energy $E_\gamma=100-200$~GeV. The angular
spread of these photons can be estimated as angular spread
of initial beam (about 0.1~mrad). The bremsstrahlung
photons have additional angular spread of about
$1/\gamma\approx 2\cdot 10^{-6}$. These two spreads are
negligible for our problem.

Then these photons collide with nuclei and produce pions,
 \be
\gamma N\to N'+\pi's,\quad \sigma\approx 110\,\mu b.
 \ee
This process gives about $10^{-3}$ $\gamma N$ collisions
per 1 electron, which is about $ 3\cdot 10^{11}$ $\gamma N$
collisions per second. On average, each of this collisions
produces one pion with high energy $E_\pi>E_\gamma/2$ (for
estimates $\la E_\pi^h\ra=70$~GeV) and at least 2-3 pions
with lower energy (for estimates, $\la E_\pi^\ell\ra\approx
20$~GeV).

Mean transverse momentum of these pions is 350-500 MeV. The
angular spread of high energy pions with energy $\la
E_\pi^h\ra$  is within 7 mrad. The increase of angular
spread of pions with decrease of energy is compensated by
growth of the number of produced pions. Therefore, for
estimates we accept that the pion flux within angular
interval 7~mrad contains $3\cdot 10^{11}$~pions with
$E_\pi=\la E_\pi^h\ra$ and the same number of pions with
$E_\pi=\la E_\pi^\ell\ra$ per second. Let us denote the
energy distribution of  pions near forward direction by
$f(E)$.

Certainly,  more refined calculations  should also consider
production and decay of $K$ mesons, etc. Reaction mentioned
in ref.~\cite{Telnov}
 \begin{subequations}\label{nutsorce}
 \be
\gamma N\to D_s^\pm X\to \nu_\tau \bar{\tau} X\,.
 \ee
plays the most essential role for our estimates. Its cross
section rapidly increases with energy growth and
 \be
\sigma\approx 2\cdot 10^{-33}\, cm^2\;\; at \;\;
E_\gamma\approx 50 \mbox{ GeV}\,.
 \ee
 \end{subequations}

\subsection{Neutrino transformer (NT). Neutrino beams}

For the neutrino transformer (NT) we suggest a low vacuum
pipe of length $L_{NT}\approx 1$~km and radius
$r_{NT}\approx 2$~m. Here muon neutrino $\nu_\mu$ and
$\bar{\nu}_\mu$ are created from $\pi\to\mu\nu$ decay. This
length $L_{NT}$ allows  more than one quarter of pions with
$E_\pi\le\la E_\pi^h\ra$ to decay. The pipe with radius
$r_{NT}$ gives an angular coverage of 2 mrad, which cuts
out 1/12 part of total flux  of low and medium energy
neutrinos. With the growth of pion energy two factors act
in opposite ways. First, with this growth initial angular
spread of pions decreases, therefore the fraction of flux
cut out by the pipe increases. Second, with this growth the
number of pion decays within relatively short pipe
decreases. These two tendencies compensate each other in
the resulting flux.

The energy distribution of neutrino's obtained from $\pi$
decay with energy $E$ is uniform in the interval $(aE,\,0)$
with $a=1-(m_\mu /m_\pi)^2$. Therefore, the energy
distribution in neutrino energy $\vep$ is obtained from
energy distribution of pions near forward direction $f(E)$
as (note that $f(E)=0$ at $E>E_e$)
 \be
F(\vep)=\int\limits_{\vep/a}^\infty f(E)dE/(aE)\,,\qquad
a=1-\fr{m_\mu^2}{m_\pi^2}\approx 0.43\,.\label{spectr}
 \ee
The increase of angular spread in the decay is negligible
in the rough approximation. Finally, at the end of NT we
expect to have the neutrino flux within the angle 2~mrad
 \bear{c}
0.6\cdot 10^{10} \nu/s \;\; with\;\; E_\nu=\la
E_\nu^h\ra\approx 30~GeV,\\ \mbox{ and }\;\; 0.6\cdot
10^{10} \nu/s \;\; with\;\; E_\nu=\la E_\nu^\ell\ra\approx
9~GeV.
 \eear{nucount}
We denote below neutrino's with $\la E_\nu\ra=30$~GeV and
$9$~GeV as {\it high energy neutrino's} and  {\it low
energy neutrino's} respectively.

\bu {\bf\bm The background $\nu_\tau$ beam}.

\bu {\bf\bm The background $\nu_\tau$ beam}.

The $\tau$ neutrino are produced  in PP. Two mechanisms
were discussed in this respect, the Bethe-Heitler process
$\gamma N\to \tau\bar{\tau}+X$ \cite{Skrinsky} and process
\eqref{nutsorce} which is is dominant \cite{Telnov}. The
cross section \eqref{nutsorce} is 5 orders less than
$\sigma(\gamma N\to X)$. Mean transverse momentum of
$\nu_\tau$ is given by $m_\tau$, which is more than 3 times
higher than that for $\nu_\mu$. Along with e.g.
$\bar{\nu}_\tau$ produced in this process, in NT each
$\tau$ decays to $\nu_\tau$ plus other particles.
Therefore,  each such reaction is a source of a
$\nu_\tau+\bar{\nu}_\tau$ pair. Finally, for flux density
we have
 \be
N_{\nu_\tau}\sim  10^4\nu_\tau/(s\cdot mrad^2)\lesssim
8\cdot 10^{-6} N_{\nu_\mu}\,.\label{nutflux}
 \ee
The $\nu_\tau$ (or $\bar{\nu}_\tau$) energy is typically
higher than that of $\nu_\mu$ by factor $2\div 2.5$.

Besides,  $\nu_\tau$ will be produced by non-decayed pions
within the protecting wall behind NP in the process like
$\pi N\to D_sX\to \tau\nu_\tau X$. The cross section of
this process increases  rapidly with energy growth and
equals $0.13\mu$b at $E_\pi=200$~GeV \cite{tel1}. Rough
estimate shows that the number of additional $\nu_\tau$
propagating in the same angular interval is close to the
estimates given in \eqref{nutflux}. In the numerical
estimates below we consider for definiteness first
contribution only. The measurement of $\nu_\tau$ flux in
the NBD is a necessary component for the study of neutrino
oscillations in FDD.

\bu Other sources of $\nu_{\mu}$ and $\nu_e$ change these
numbers only weakly.

\subsection{ Nearby detector (NBD)}

\bu {\bf Main goal} of nearby detector (NBD) is to measure
the energy and angular distribution of neutrino's within
the beam as well as $N_{\nu_e}/N_{\nu_\mu}$ and
$N_{\nu_\tau}/N_{\nu_\mu}$.

\bu We suggest to position the NBD behind NT and a concrete
wall (to eliminate pions and other particles from initial
beam).  For estimates, we consider the NBD in a form of
water cylinder of radius about 2-3~m (roughly the same as
NT) and length $\ell_{NBD}\approx 100$~m.

For $E_\nu=30$ GeV the cross section for $\nu$ absorbtion
is
 \bear{c}
 \sigma(\bar{\nu} N\to \mu^+h)=0.1\fr{m_pE_{\bar{\nu}}}{\pi v^4}
 \approx
10^{-37} cm^2,\\ \sigma(\nu N\to
\mu^-h)=0.22\fr{m_pE_\nu}{\pi v^4}\approx 2\cdot 10^{-37}
cm^2.
 \eear{}
At these numbers  the free path length in water is
 $\lambda_{\bar{\nu}}=10^{13}$~cm and $\lambda_\nu= 0.45\cdot
10^{13}$~cm. That gives
 \bear{c}
(3\div 6)\cdot 10^7\;\; \mu/{\mathbf {year}}\;\;
 (with\;\;\la E_\mu\ra\sim 30\;GeV);\\
400\div 800\;\; \tau/{\mathbf {year}}\;\;  (with\;\;\la
E_\tau\ra\sim 50\;GeV)
 \eear{numberNBD}
(here 1 year =$10^7$~s). These numbers look sufficient for
detailed measurements of muon neutrino spectra and
verification of calculations of $\nu_\tau$ backgrounds.

\subsection{Far Distance Detector (FDD)}

\bu {\it Main goal of FDD} --- {\bf study of neutrino
oscillations.} We consider $\nu_\mu-\nu_\tau$ oscillations
and oscillations with sterile neutrino. We consider
separately the potentials of two possible positions of FDD,
assuming  the length of oscillations to be \cite{Vysot}
 \be
 L_{osc}\approx
E_\nu/(50~GeV)\cdot 10^5~km\,.\label{Losc}
 \ee

\bu {\bf FDD I}

The first opportunity is to place FDD at the distance
$L_F=200$~km. In this case the initial beam should be
turned at 16~mrad  angle. This angle can be reduced by
3~mrad (one half of angular spread of initial pion beam).

We consider this FDD in the form of water channel of length
1~km with radius $R_F\approx 40$~m. The transverse size is
limited by water transparency.

The fraction of neutrino's reaching this FDD is given by
ratio\linebreak[4] $k=(R_F/L_F)^2/[(r_{NT}/L_{NT})^2]$. In
our case $k=0.01$. Main effect under interest here is
$\nu_\mu\to \nu_\tau$ oscillation.  They add
$(L_F/L_{osc})^2N_{\nu_\mu}$ to initial $N_{\nu_\tau}$.

In FDD of chosen sizes we expect the counting rate to be
just 10 times lower than that in NBD \eqref{numberNBD} for
$\nu N\to \mu X$ reactions with high energy neutrino. We
also expect the rate of $\nu_\tau N\to \tau X$ events to be
another $10^5$ times lower (that is about 10 times higher
than the background given by initial $\nu_\tau$ flux,
  \be
\begin{array}{c}
  N(\nu_\mu N\to \mu X)\approx (3\div 6)\cdot
  10^6/year,\\
N(\nu_\tau N\to \tau X)\approx (30\div
60)/year\end{array}\;\; in\;\; FDDI.\label{FDDInumb}
 \ee
For neutrino of lower energies effect  increases. Indeed,
$\sigma(\nu N\to\tau X)\propto E_\nu$ while $L_{osc}\propto
E_\nu$. Therefore,  observed number of $\tau$ from
oscillations increases $\propto 1/E_\nu$ at $E_\nu\ge
10$~GeV. The additional counting rate for $\nu_\tau N\to
\tau X$ reaction with low energy neutrino (with $\la
E_\nu\ra=9$~GeV) cannot be estimated so simply, but rough
estimates give   numbers similar to \eqref{FDDInumb}.

These numbers look sufficient for separation of
$\nu_\mu-\nu_\tau$ oscillations and rough measurement of
$s_{23}$.

Note that at given FDD size the counting rate of $\nu_\tau
N\to \tau X$ reaction is independent on FDD distance from
LC, $L_F$. The growth of $L_F$ improves the signal to
background ratio for oscillations. The value of signal
naturally increases with growth of volume of
FDD.\\

\bu {\bf FDD II}

The second opportunity is to use for FDD well known {\it
Ice-cub detector} in Antarctica with volume 1 km$^3$. The
distance to FDD in this case is $L_F\approx 10^4$~km. This
opportunity requires relatively expensive excavation work
for NT and NBD at the angle about $60\deg$ under horizon.

At this $L_F$ for $\nu$ with energy about 30~GeV we expect
the conversion of $(L_F/L_{osc})^2\approx 1/36$ for
$\nu_\mu\to \nu_\tau$ or $\nu_\mu\to sterile\; \nu$.

In this FDD the number of expected events $\nu_\mu\to \mu
X$ with high energy neutrino will be about 0.01 of that in
NBD,
 \be
 \begin{array}{c}
  N(\nu_\mu N\to \mu X)\approx (3\div 6)\cdot
  10^5/year,\\
N(\nu_\tau N\to \tau X)\approx 10^4/year\end{array} \;\;
in\;\; FDDII. \label{FDDIInumb}
 \ee
The contribution of low energy neutrino increases both
these counting rates.

Therefore, one can hope that a few years of experimentation
with reasonable $\tau$ detection efficiency will allow to
measure $s_{23}$ with percent accuracy, and similar period
of observation of $\mu$ production will allow to observe
the loss of $\nu_\mu$ due to transition this neutrino to
sterile $\nu$.

\section{Discussion }

\bu All technical details of proposed scheme including
sizes of all elements, construction, and materials of
detectors can be modified in the forthcoming simulations
and optimization of parameters. The numbers obtained above
represent first rough estimates only. In particular, we did
not discuss here methods of $\mu$ and $\tau$ registration
and their efficiency. Next,  large fraction of residual
electrons, photons and pions leaving the PP will reach the
walls of the NT pipe. The  heat sink and radiation
protection of this pipe must be taken into account.

\bu More detailed physical program of this neutrino factory
will be similar to the one discussed in other projects
\cite{nufact,nufact1}.

\section{Other possible applications of LC used beam }

\bu {\bf Pion producer of neutrino factory in the fixed
target experiment}. The proposed PP can be used also as
$\gamma N$ collider with luminosity $3\cdot
10^{39}$~cm$^{-2}$s$^{-1}$. Therefore, the PP with
additional standard detector behind PP can be used for
precise experiments in the fixed target regime for the
$\gamma N$ collider with huge luminosity. Here one can
study  rare processes in $\gamma N$ collisions, $B$
physics, etc.

\bu {\bf Additional opportunity for using NBD of neutrino
factory.} High rate of $\nu_\mu N\to \mu X$ processes
expected in NBD allows to study  new problems of high
energy physics. The simplest example is the opportunity to
study  charged and axial current induced diffraction ($\nu
N\to\mu N' \rho^\pm N$, $\nu N\to\mu N' b_1^\pm N$,...)
with high precision. Measurements of charged current
induced structure functions present the second example.

\bu {\bf Material sciences}. The  interaction of beam
having exceptional energy density \eqref{beampar} with
different materials will be of great  interest for material
physics (for example, to understand what happens at
collision of micro-meteorite with  spacecraft).\\

I am thankful to  D.~Naumov, L.~Okun, V.~Saveliev,
A.~Sessler, A.~Skrinsky, V.~Telnov, M.~Vysotsky,
M.~Zolotarev for comments and new information. This work is
supported by grants RFBR 05-02-16211, NSh-2339.2003.2.

\end{document}